\documentclass[10pt,conference]{IEEEtran}
\usepackage{graphicx,subfigure,psfrag,amsmath,cases}
\usepackage{latexsym,amsmath,amssymb,subfigure,algorithm,mathtools}
\usepackage{algorithmic}
\usepackage{color}
\usepackage{url}
\usepackage{scrtime}
\usepackage{cite}
\usepackage{subfigure}
\usepackage{bbding}
\usepackage{multicol}
\usepackage{bm}

\author{
{Ruide Li}\IEEEauthorrefmark{1}\IEEEauthorrefmark{2}, {Zhiqiang Wei}\IEEEauthorrefmark{2}, {Lei Yang}\IEEEauthorrefmark{2}, {Derrick Wing Kwan Ng}\IEEEauthorrefmark{2}, {Nan Yang}\IEEEauthorrefmark{3}, {Jinhong Yuan}\IEEEauthorrefmark{2}, and {Jianping An}\IEEEauthorrefmark{1}\\
\IEEEauthorblockA{\IEEEauthorrefmark{1}{School of Information and Electronics, Beijing Institute of Technology, Beijing, China}\\\IEEEauthorrefmark{2}{School of Electrical Engineering and Telecommunications, The University of New South Wales, Sydney, Australia} \\ \IEEEauthorrefmark{3}{Research School of Engineering, The Australian National University, Canberra, Australia}\\
Email: taiyuanlaide@163.com, zhiqiang.wei@student.unsw.edu.au,  \{lei.yang3, w.k.ng\}@unsw.edu.au,\\
yangnan1616@gmail.com, j.yuan@unsw.edu.au, an@bit.edu.cn}
\vspace{-10mm}
}

\title{Joint Trajectory and Resource Allocation Design for UAV Communication Systems}

\newtheorem{T-Prob}{Transformed Problem}
\newtheorem{proposition}{Proposition}

\DeclareMathOperator{\maxo}{maximize}

\begin{document}
\maketitle
\begin{abstract}
  In this paper, we investigate resource allocation design for unmanned aerial vehicle (UAV)-enabled communication systems, where a UAV is dispatched to provide communications to multiple user nodes.
  Our objective is to maximize the communication system throughput by jointly optimizing the subcarrier allocation policy and the trajectory of the UAV, while taking into account the minimum required data rate for each user node, no-fly zones (NFZs), the maximum UAV cruising speed, and initial/final UAV locations.
  The design is formulated as a mixed integer non-convex optimization problem which is generally intractable.
  Subsequently, a computationally-efficient iterative algorithm is proposed to obtain a locally optimal solution.
  Simulation results illustrate that the performance of the proposed iterative algorithm
  approaches closely to that of the system without NFZ.
  In addition, the proposed algorithm can achieve a significant throughput gain compared to various benchmark schemes.\vspace*{-1.5mm}
\end{abstract}

\section{Introduction}\vspace*{-1.5mm}
The rapid growing demand on wireless communication services, e.g. ultra-high data rates and massive connectivity \cite{wong_2017_Key_Technologies_for_5G}, has fueled the development of wireless networks and the mass productions of wireless devices.
Despite the fruitful research in the literature for providing ubiquitous services, the performance of wireless systems is limited by users with poor channel condition.
Fortunately, owing to the high flexibility and low cost in deployment of unmanned aerial vehicles (UAVs), UAV-enabled communication offers a promising solution to tackle this challenge.
In particular, the high mobility of UAVs facilitates the establishment of strong line-of-sight (LoS) links to ground users.
Hence, in recent years, numerous applications of UAV-enabled communication have emerged dramatically not only in the military domain, but also in the civilian and commercial domains, such as disaster relief, archeology, pollution monitoring, commodity deliver, etc. \cite{Zeng_2016_Unmanned_Aerial_Vehicles_Magazine}.
Besides, several world leading industrial companies, such as Facebook, Google, and Qualcomm, have made advancements on their journey to deliver high-speed internet from the air by UAVs.
Furthermore, the utilization of UAVs for wireless networks has recently received significant attention from the academia, such as mobile relays \cite{Zeng_2016_Throughput_Maximization_for_UAV}, aerial mobile base stations \cite{Device_to_Device_Communications_Mozaffari}, and UAV-enabled information dissemination and data collection \cite{Zeng_2017_Completion_Time_Minimization_in_UAV}.

UAV-enabled information dissemination or multicasting is one of the most important applications.
To provide communication services to multiple downlink users, \cite{Resource_Allocation_for_Solar_Powered} studied a multiple access scheme with a solar-powered UAV.
%
Yet, \cite{Resource_Allocation_for_Solar_Powered} only considered the system throughput maximization without taking into account the quality of services (QoS) requirement in communications.
\cite{zhang2017cellular} studied the mission completion time minimization of a cellular-enabled UAV communication system which guarantees the QoS of connectivity between the UAV and ground base stations.
However, utilizing UAV to provide wireless communications to multiple users while ensuring the minimum data rate required for each user is not considered in \cite{Resource_Allocation_for_Solar_Powered}, and \cite{zhang2017cellular}.

Recently, trajectory design or path planning has been a major research area in the existing literature on UAV-enabled communications.
For example, \cite{Zeng_2016_Throughput_Maximization_for_UAV} optimized the trajectory of a UAV to maximize the system throughput of a single-user while taking into account its maximum mobility.
Authors in \cite{Zhang_2017_Securing_UAV_Communications} investigated the UAV's trajectory design to guarantee secure air-to-ground communications.
Although the trajectory design in the literature has focused on different scenarios, they do not consider the geometrical restrictions in UAV trajectory design.
For example, due to regulations for military, security, safety or privacy reasons, there are some no-fly zones (NFZs) which flight of UAVs is prohibited \cite{Valavanis:2014:HUA:2692452}, \cite{zhao_chen_yu_2017}.
To the best of our knowledge, only limited research efforts have been investigated in the literature.
For instance, the authors of \cite{Valavanis:2014:HUA:2692452} proposed a control-mechanism based on the geometrical tangential method of control theory to avoid the UAV flying into the NFZ.
However, their proposed method only focused on the constraint of NFZ and neglected the air-to-ground communication for satisfying the data rate requirements of user nodes.

In this paper, we consider a rotary-wing UAV-enabled orthogonal frequency division multiple access (OFDMA) communication system, where a UAV is dispatched to provide communications to multiple ground user nodes with guaranteed QoS requirements.
Meanwhile, the UAV is not allowed to fly over the NFZs.
We aim to maximize the communication system throughput by jointly designing the subcarrier allocation policy and the UAV's trajectory, subject to the per-user minimum data rate requirement, the existence of NFZs, the UAV's maximum speed, and the initial/final locations.
Simulation results demonstrate that the performance of our proposed algorithm approaches that of the system without NFZ.\vspace*{-1mm}

\begin{figure}[th]
  \centering
  \includegraphics[width=2.5 in]{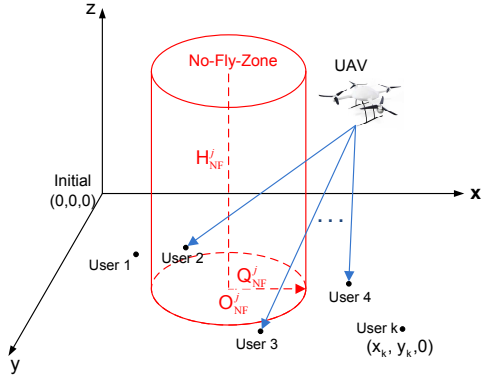}\vspace*{-2mm}
  \caption{A system model of a UAV-enabled communication system with the restriction of an no-fly zone.}\vspace*{-6mm}
  \label{fig:UAV communication frame}
\end{figure}

\section{System Model and Problem Formulation}\vspace*{-1mm}
We consider a UAV-enabled wireless communication system consisting of one rotary-wing UAV and $K$ downlink user nodes denoted by the set $k\in\mathcal{K} \triangleq \{1, \cdots, K\} $, as shown in Fig. \ref{fig:UAV communication frame}.
OFDMA is adopted at the UAV to provide communications to user nodes.
The system bandwidth is divided equally into $N_\mathrm{F}$ orthogonal subcarriers.
We express the locations of all nodes in a three-dimensional (3D) Cartesian coordinate system.

Denoting the flight period of the UAV as $T$ in seconds (s).
For the ease of design, the time $T$ is discretized into $N$ time slots with equal-length, $\delta t$, $0\le t\le T$, which is small enough to denote the distance between the UAV and the user node as a constant within each time slot.\footnote{The discretized model is commonly adopted in the literature to facilitate the design of resource allocation for UAV-enabled communication systems \cite{Zeng_2016_Throughput_Maximization_for_UAV}, \cite{Zhang_2017_Securing_UAV_Communications}.}
Thus, the ground projected trajectory of the UAV, $(x(t),y(t))$, over the time $T$ can be approximated by the sequence $\{x[n],y[n]\}_{n=1}^N$, where $\{x[n],y[n]\} \triangleq \{x(n\delta t), y(n\delta t)\}$, $\forall n$, denotes the horizontal location of the UAV at time slot $n$.
Denote the maximum speed of UAV as $v_{\rm{max}}$ in meters per second (m/s), and the UAV's maximum aviation distance in each time slot is $V=v_{\rm{max}}\delta t$ in meters (m).
In particular, we assume that the UAV's initial location
and the final location projected on the ground is $\{x[0],y[0]\}\triangleq{(0,0)}$ and $\{x[N],y[N]\}$, and user node $k$ is located at $(x_k, y_k, 0)$, $\forall k$, without lost of generality.
Denote $d_k[n]$, $\forall n,k$, as the distance between the UAV and user $k$ at time slot $n$, which can be written as:\vspace{-1.5mm}
\begin{equation}\label{eqn:distance of communication}
  d_k^2[n] = (x[n] - x_k)^2 + (y[n] - y_k)^2 + H^2,\; \forall n,k,
\end{equation}
where $H$ in meters is the constant flying altitude of the UAV.

On the other hand, NFZ is an important issue that should be considered in UAV trajectory planing, as a UAV is strictly prohibited to fly over the NFZs.
In this paper, we assume that there are ${N_\mathrm{NF}}$ non-overlapped NFZs.
Specifically, we define NFZ $j\in\{1,...,N_\mathrm{NF}\}$, as a cylindrical region with a coordinate center $(x_\mathrm{NF}^j, y_\mathrm{NF}^j)$ projected on the ground, height $H_\mathrm{NF}^j$, $H<H_\mathrm{NF}^j$, $\forall j$, and radius $Q_\mathrm{NF}^{j}$, $\forall j$, cf. Fig. \ref{fig:UAV communication frame}.
With the existence of NFZs, the trajectory of the UAV should satisfy:\vspace{-1.5mm}
\begin{equation}\label{eqn:No-Fly area region}
  (x[n] - x_\mathrm{NF}^j)^2 + (y[n] - y_\mathrm{NF}^j)^2 \geq Q_\mathrm{NF}^{j2},\; \forall n,j.
\end{equation}

We assume that the wireless channels from the UAV to user $k$ are LoS-dominated, and thus we adopt the free-space path loss model as in \cite{Zeng_2016_Unmanned_Aerial_Vehicles_Magazine}.
Denote $s_k^i[n]\in \{0,1\}$ as the binary subcarrier allocation variable of the $i$-th, $i\in \{1,...,N_F\}$, subcarrier for user $k$ at time slot $n$. We have $s_k^i[n]=1$ if user $k$ is allocated to subcarrier $i$, at time slot $n$ and, $s_k^i[n]=0$, otherwise.
Furthermore, each subcarrier can be allocated to at most one user to avoid multiple access interference. Thus, with \eqref{eqn:distance of communication}, the communication rate $R_k[n]$ from the UAV to user $k$ in bits/second/Hz (bps/Hz) at time slot $n$ over all subcarriers can be written as\vspace*{-2mm}
\begin{equation}\label{eqn:achievable rate from U2k}
  R_k[n] = \sum_{i=1}^{N_\mathrm{F}}s_k^i[n]\log_2 \bigg (1 + \frac{\gamma_0 \mathcal{P}}{d_k^2[n]} \bigg), \; \forall k, n,\vspace*{-2mm}
\end{equation}
\noindent where $\gamma_0 = \beta_0/ \sigma^2$ represents the reference signal-to-noise ratio (SNR) and $\beta_0$ denotes the channel power gain at the reference distance $d_0 = 1$ m, which depends on the signal-carrier frequency, antenna gain, etc.
Besides, $\sigma^2$ denotes the additive white Gaussian noise (AWGN) power at ground user nodes. Also, to reduce the peak-to-average power ratio in the considered multicarrier systems, we assume that the transmit power\footnote{In this work, we assume a fixed power allocation to simplify the resource allocation design. Adaptive power allocation will be considered in the future work.} on each subcarrier at the UAV is $\mathcal{P}$.

The joint trajectory and subcarrier allocation design can be formulated as the following optimization problem:\vspace*{-1mm}
\begin{align}\label{eqn:problem 1}
\underset{\bold{x},\bold{y}, \bold{s}}{\maxo} &\sum_{n=1}^{N}\sum_{k=1}^{K}\sum_{i=1}^{N_\mathrm{F}} s_k^i[n] \log_2\left(1 + \frac{\gamma_0\mathcal{P}}{d_k^2[n]}\right) \\
  \mathrm{s.t.}\,\mathrm{C1}: &\sum_{i=1}^{N_\mathrm{F}}s_k^i[n]\log_2 \bigg (1 + \frac{\gamma_0 \mathcal{P}}{d_k^2[n]} \bigg) \geq R^k_\mathrm{min},\forall n,k,\notag\\
  \mathrm{C2}: &\sum_{k=1}^{K}s_k^i[n]\leq1, \forall n, i, \notag\\
  \mathrm{C3}: & s_k^i[n]\in \{0,1\}, \forall n, k, i, \notag\\
  \mathrm{C4}: &(x[n] - x_\mathrm{NF}^j)^2 + (y[n] - y_\mathrm{NF}^j)^2 \geq (Q_\mathrm{NF}^{j})^2, \forall n, j, \notag\\
  \mathrm{C5}: &(x[n]-x[n-1])^2 + (y[n]-y[n-1])^2 \leq V^2, \forall n,\notag
  \end{align}
\noindent where $\bold{x} = \{x[n], \forall n\}$, $\bold{y} = \{ y[n], \forall n\}$, and $\bold{s} = \{s_k^i[n], \forall k, n, i\}$ are the sets of the optimization variables. Constraint $\mathrm{C1}$ is imposed to guarantee a minimum required data rate $R^k_\mathrm{min}$ for user $k$, $\forall k$, at each time slot.
Constraints $\mathrm{C2}$ and $\mathrm{C3}$ are introduced to ensure that each subcarrier can be allocated to at most one user.
Constraint $\mathrm{C4}$ is the NFZ constraint such that a UAV is strictly prohibited to fly over the NFZs.
Constraint $\mathrm{C5}$ is the UAV's maximum movement speed constraint.

Problem \eqref{eqn:problem 1} is a mixed-integer non-convex optimization problem where the non-convexity arises from the objective function, constraints $\mathrm{C1}$, $\mathrm{C3}$, and $\mathrm{C4}$.
Therefore, it is challenging to solve the problem \eqref{eqn:problem 1} optimally with a low computational complexity.\vspace*{-1.5mm}

\section{Joint Trajectory and Resource Allocation Design}\vspace*{-1.5mm}
In this section, we divide problem \eqref{eqn:problem 1} into two subproblems, and propose a computationally efficient iterative alternating algorithm to achieve a suboptimal solution.
In particular, we first study the optimal resource allocation for a given UAV trajectory, which is denoted by subproblem 1.
Then, for a given resource allocation policy, we optimize the UAV's trajectory.
\vspace*{-1.5mm}

\subsection{Subproblem 1: Resource Allocation Optimization}\vspace*{-1.5mm}

In this section, we consider subproblem 1 for optimizing the resource allocation by assuming that the UAV's trajectory $\{x[n],y[n]\}_{n=1}^N$ is fixed.
Thus, subproblem 1 can be written as:\vspace*{-1mm}
\begin{align}\label{eqn:problem_1.1}
\underset{\bold{s}}{\maxo}\,\,
&\sum_{n=1}^{N}\sum_{k=1}^{K}\sum_{i=1}^{N_\mathrm{F}} s_k^i[n] \log_2\left(1 + \frac{\gamma_0\mathcal{P}}{d_k^2[n]}\right) \\
  \mathrm{s.t.}\,\mathrm{C1}:\,\, &\sum_{i=1}^{N_\mathrm{F}}s_k^i[n]\log_2 \bigg (1 + \frac{\gamma_0 \mathcal{P}}{d_k^2[n]} \bigg) \geq R^k_\mathrm{min},\forall n,k,\notag\\
  \mathrm{C2}:\, &\sum_{k=1}^{K}s_k^i[n]\leq1, \forall n, i, \notag\\
  \mathrm{C3}:\, &s_k^i[n] \in \{ 0,1\} ,\forall n, k, i.\notag
\end{align}
\vspace*{-5mm}

The optimization problem in \eqref{eqn:problem_1.1} is an integer linear programming problem \cite{A_Multiplier_Adjustment_Method}, which is challenging to solve with a low computation complexity.
To circumvent the difficulty, we first relax the binary constraint C3 in \eqref{eqn:problem_1.1} as $\overline{\rm{C3}}$: $0<s_k^i[n]\le1$, $\forall n, k, i$, and study its solution structure. Then, the obtained solution from the constraint relaxed problem will serve as a building block for the development of the optimal solution of the original problem.
After replacing C3 with $\overline{\rm{C3}}$ in \eqref{eqn:problem_1.1}, the associated Lagrangian function is given by:\vspace*{-1mm}
\begin{align}\label{eqn:problem_1.1_Lag}
{\cal L}({\bm{\alpha }},{\bm{\beta }},{\bf{s}})
 = &\sum\limits_{n = 1}^N \sum\limits_{k = 1}^K \sum\limits_{i = 1}^{{N_\mathrm{F}}}\left( \left( {{\alpha _{n,k}} + 1} \right){R_k}[n]    -  {\beta _{n,i}}\right) s_k^i[n]   \notag\\
 - & {}\sum\limits_{n = 1}^N\sum\limits_{k = 1}^K {{\alpha _{n,k}}R^k_{{\rm{min}}}}  + \sum\limits_{n = 1}^N {\sum\limits_{i = 1}^{{N_{\rm{F}}}} {{\beta _{n,i}}} }, \,\,\forall n,k,i,
\end{align}
\vspace*{-3mm}

\noindent where ${\bm{\alpha}} \triangleq \{\alpha_{n,k}, \forall n,k\} \ge 0$, and ${\bm{\beta} \triangleq \{\beta_{n,i}, \forall n,i\}} \ge 0$ are the Lagrange multipliers corresponding to constraints $\mathrm{C1}$ and $\mathrm{C2}$, respectively.
Then, the Lagrange dual function can be defined as the maximum value of the Lagrangian over ${\bf{s}}$, which is obtained by \cite{Boyd}:\vspace*{-1.5mm}
\begin{equation}\label{eqn:problem 1.1_Dual}
g({\bm{\alpha}},{\bm{\beta}})= \mathop {\sup }\limits_{\bf{s}}\;\; {\cal L}({\bm{\alpha }},{\bm{\beta }},{\bf{s}}).\vspace*{-1.5mm}
\end{equation}

We can observe that the Lagrange dual function in \eqref{eqn:problem_1.1_Lag} is bounded if and only if\vspace*{-1.5mm}
\begin{equation}\label{eqn:problem_1.1_Lag_s_i}
\left( {{\alpha _{n,k}} + 1} \right){R_k}[n] - {\beta _{n,i}} = 0, \,\, \forall n,k,i.\vspace*{-1.5mm}
\end{equation}
Hence, the dual problem of \eqref{eqn:problem_1.1} with relaxed constraint $\overline{\rm{C3}}$ is given by:
\begin{align}\label{eqn:problem_1.1_Lag_relax}
\mathop {\inf }\limits_{{\bm{\alpha }},{\bm{\beta }}\ge 0} &\mathop {\sup }\limits_{\bf{s}}\;\; {\cal L}({\bm{\alpha }},{\bm{\beta }},{\bf{s}}),  \\
\mathrm{s.t.} \;\; &\eqref{eqn:problem_1.1_Lag_s_i}.\notag \vspace*{-1mm}
\end{align}

\begin{table}[t]\label{table:algorithm 1}\vspace*{-2mm}
  \begin{algorithm} [H]                    
  \renewcommand\thealgorithm{1}
  \caption{Optimal Resource Allocation}
  \label{algorithm:algorithm 1}     
   \begin{algorithmic} [1]
   \STATE For a user set $k\in\mathcal{K} \triangleq \{1, \cdots, K\}$ and a given UAV trajectory.

   \FOR{$n=1:N$}
   \STATE Find the strongest user $k^* = \underset{\mathcal{K}}{\maxo}\; R_k[n]$ at time slot $n$.

   \STATE Compute the minimum number of subcarriers for all users according to $N_k[n]=\lceil {\frac{R^k_{{\rm{min}}}}{R_k[n]}} \rceil$, $\forall k\in\{\mathcal{K}/k^*\}$,

   \STATE Compute the number of subcarriers for the strongest user by: $N_{k^*}[n]=N_F - \sum\limits_{k\in\{\mathcal{K}/k^*\}}^{}N_k[n]$

   \IF {$N_{k^*}[n]R_{k^*}[n] \ge R^k_{\rm{min}}$}
   \STATE Set $i=1$.
   \STATE Allocate $N_{k}[n]$ subcarriers to user $k$
   \FOR{$k=1:K$}
   \FOR {$i_\mathrm{index}=1: N_k[n]$}
   \STATE $s_k^i[n] = 1$.
   \STATE $i = i+1$.
   \ENDFOR
   \ENDFOR
   \ELSE
   \STATE Stop and declare problem \eqref{eqn:problem_1.1} infeasible.
   \ENDIF
  \ENDFOR

   \end{algorithmic}
  \end{algorithm}\vspace*{-11mm}
\end{table}

\vspace*{-2mm}
Now, the constraints in the dual problem only holds when ${\beta _{n,i}} = {\beta _{n,j}} >0$, $\forall n,i,j$.
It means that all the subcarriers share the same value of the Lagrange multiplier for constraint C2.
In fact, the channel from the UAV to user node $k$ are LoS-dominated \cite{Zeng_2016_Unmanned_Aerial_Vehicles_Magazine}, which leads to a frequency-flat fading across all the subcarriers.
The second important observation is that, with ${R_k}[n] \neq {R_{k'}}[n]$, $\forall n,k,k'$, at most one ${\alpha _{n,k^{*}}}$ can be zero with ${k^*} = \mathop {\max }\limits_k {R_k}[n]$ at each time slot $n$.
In particular, if ${\alpha _{n,k}} = {\alpha _{n,k'}} = 0$, $\forall n$, we have ${R_k}[n] = {R_{k'}}[n] = {\beta _{n,i}}$ based on \eqref{eqn:problem_1.1_Lag_s_i}, which leads to a contradiction.
On the other hand, if ${R_k}[n] > {R_{k'}}[n]$ with ${\alpha _{n,k}} >0$ and ${\alpha _{n,k'}} = 0$, we have $\left( {{\alpha _{n,k}} + 1} \right){R_k}[n] = {R_{k'}}[n] = {\beta _{n,i}}$, which also leads to a contradiction.
Therefore, the Lagrange multiplier of the minimum rate constraint C1 for the strongest user ${k^*} = \mathop {\max }\limits_k {R_k}[n]$ is the only one to be zero at the optimal point.\footnote{For the special case with more than one users having identical largest rate among all the users at time slot $n$, i.e., $[k,k'] = \mathop {\max }\limits_k {R_k}[n]$, it can permit more than one ${\alpha _{n,k}}$ to be zero, i.e., ${\alpha _{n,k}} = {\alpha _{n,k'}} = 0$.}
The physical insight of this observation can be revealed based on the Karush-Kuhn-Tucker (KKT) conditions.

According to the KKT conditions \cite{Boyd}, the following equation should hold at the optimal point of the problem in \eqref{eqn:problem_1.1} with the relaxed constraint $\overline{\rm{C3}}$:\vspace*{-1.5mm}
\begin{equation}\label{eqn:problem 1.1_Dual}
{\alpha _{n,k}}\left( {\sum\limits_{i = 1}^{{N_{\rm{F}}}} {s_k^i} \left[ n \right]{R_k}\left[ n \right] - {R^k_{{\rm{min}}}}} \right) = 0, \forall n,k.\vspace*{-1.5mm}
\end{equation}
Therefore, if ${\alpha _{n,k}} > 0$, the minimum data rate requirement C1 of user $k$ at time slot $n$ must be satisfied with equality.
On the other hand, if ${\alpha _{n,k}} = 0$, i.e., user $k$ is the strongest user at time slot $n$, we have ${\sum\nolimits_{i = 1}^{{N_{\rm{F}}}} {s_k^i} \left[ n \right]{R_k}\left[ n \right] \ge {R^k_{{\rm{min}}}}}$.
The physical meaning of this observation is that no exceedingly large amount of resources should be allocated to any users except for the strongest user $k^*$ when user $k$, $\forall k\in\{\mathcal{K}/k^*\}$, such that user $k$ satisfies constraint C1 with equality.

\underline{\textit{Optimal Subcarrier Allocation}:}
Now, due to the binary constraint $\mathrm{C3}$ in \eqref{eqn:problem_1.1}, the minimum data rate requirement C1 of user $k$ at time slot $n$ may not hold with equality at the optimal point.
However, based on the insight from \eqref{eqn:problem_1.1_Lag_relax} and \eqref{eqn:problem 1.1_Dual}, we can propose a sequentially subcarrier allocation algorithm summarized in {\bf Algorithm \ref{algorithm:algorithm 1}}.
Note that $\lceil \cdot \rceil$ is the ceiling function which returns the smallest integer greater than the input value.

\begin{proposition}
The proposed algorithm can achieve the optimal solution of problem \eqref{eqn:problem_1.1}. In particular, except the strongest user, other users are allocated with minimum number of subcarriers to satisfy constraint C1. Then, the  remaining subcarriers should be allocated to the strongest users to achieve the optimal solution of \eqref{eqn:problem_1.1}.
\end{proposition}

\bm{{\it{Proof:}}} Due to the page limitation, we only provide a sketch of proof. Assuming the optimal solution of \eqref{eqn:problem_1.1} as $N^*_k[n]$, $\forall k,n$, which denotes the number of subcarriers allocated to user $k$ at time slot $n$.
For any weaker user $k$, $\forall k\in\{\mathcal{K}/k^*\}$, reducing the number of subcarriers allocated to user $k$ will violate constraint C1, which makes problem \eqref{eqn:problem_1.1} infeasible.
On the other hand, moving $i'>0$ subcarriers from the strongest user $k^*$ to any weaker user $k$, $\forall k\in\{\mathcal{K}/k^*\}$, the system throughput will be degraded by ${i'({R_{k^*}}\left[ n \right] - {R_k}\left[ n \right]} )>0$.
This completes the proof.

Note that the order of subcarriers allocation for the weaker users will not effect the subcarrier allocation policy.
It is because the channels are frequency flat across different subcarriers. \vspace*{-1.5mm}

\subsection{Subproblem 2: Trajectory Optimization}\vspace*{-1.5mm}

In this section, we consider subproblem 2 for optimizing UAV trajectory by assuming that the resource allocation is fixed, which yields:
\vspace{-1mm}
\begin{align}\label{eqn:subproblem 2.1}
\underset{\bold{x},\bold{y}}{\maxo} &\sum_{n=1}^{N}\sum_{k=1}^{K}\sum_{i=1}^{N_\mathrm{F}} s_k^i[n] \log_2\left(1 + \frac{\gamma_0\mathcal{P}}{d_k^2[n]}\right) \\
  \mathrm{s.t.}\,\mathrm{C1}: &\sum_{i=1}^{N_\mathrm{F}}s_k^i[n]\log_2 \bigg (1 + \frac{\gamma_0 \mathcal{P}}{d_k^2[n]} \bigg) \geq R^k_\mathrm{min},\forall n,k,\notag\\
  \mathrm{C4}: &(x[n] - x_\mathrm{NF}^j)^2 + (y[n] - y_\mathrm{NF}^j)^2 \geq Q_\mathrm{NF}^{j2}, \forall n, j, \notag\\
  \mathrm{C5}: &(x[n]-x[n-1])^2 + (y[n]-y[n-1])^2 \leq V^2,\forall n.\notag
\end{align}

However, problem \eqref{eqn:subproblem 2.1} is still non-convex due to the non-convex objective function and non-convex constraints $\mathrm{C1}$ and $\mathrm{C4}$.
Although it is hard to solve the non-convex problem \eqref{eqn:subproblem 2.1} optimally, by utilizing the difference of convex (D.C.) programming, we can obtain a suboptimal solution for problem \eqref{eqn:subproblem 2.1} with a polynomial time computational complexity \cite{Optimal_Joint_Power_and_Subcarrier_Allocation}, \cite{wei2016power}.

By introducing slack variables $\bold{t} = \{t_k[n], \forall n, k\}$, the problem \eqref{eqn:subproblem 2.1} can be written as:
\begin{align}\label{eqn:subproblem 2.2}
\underset{\bold{x},\bold{y},\bold{t}}{\maxo}\,\,
 &\sum_{n=1}^{N}\sum_{k=1}^{K}\sum_{i=1}^{N_\mathrm{F}} s_k^i[n] \log_2\left(1 + \frac{\gamma_0\mathcal{P}}{t_k[n]}\right) \\
  \mathrm{s.t.}\,\mathrm{C1}: &\sum_{i=1}^{N_\mathrm{F}}s_k^i[n]\log_2 \bigg (1 + \frac{\gamma_0 \mathcal{P}}{t_k[n]} \bigg) \geq R^k_\mathrm{min},\forall n,k,\notag\\
  \mathrm{C6}: &d_k^2[n] \le t_k[n],\forall k, n,\notag\\
 \mathrm{C4},\, \, &\mathrm{C5}, \notag
\end{align}
\noindent where we introduce a new constraint $\mathrm{C6}$. It is easy to prove that, problem \eqref{eqn:subproblem 2.2} is equivalent to problem \eqref{eqn:subproblem 2.1}, since $\mathrm{C6}$ holds with equality at the optimal point of problem \eqref{eqn:subproblem 2.2}. In particular, assuming the optimal solution of problem \eqref{eqn:subproblem 2.2} as $(\bold{x}^*,\bold{y}^*,\bold{t}^*)$, if there exists a $t^*_k[n] > d_k^{2}[n] = (x^*[n] - x_k)^{2} + (y^*[n] - y_k)^{2} + H^2$, we can further improve the objective value by reducing $t^*_k[n]$ without violating constraint $\mathrm{C1}$. Therefore, it leads to a contradiction that $t^*_k[n]$ is the optimal solution.

Furthermore, for sufficiently large values $\lambda \gg 1$ and $\eta \gg 1$, constraints $\mathrm{C1}$ and $\mathrm{C4}$ can be augmented into the objective function \cite{Optimal_Joint_Power_and_Subcarrier_Allocation}, \cite{wei2016power}.
Then, we can rewrite the maximization problem \eqref{eqn:subproblem 2.2} in the canonical form of D.C. programming as follows:\vspace*{-2mm}
\begin{align}\label{eqn:P2_DCD}
\underset{\bold{x},\bold{y},\bold{t}}{\maxo}\,\, &F(\bold{x},\bold{y},\bold{t}) - G(\bold{x},\bold{y},\bold{t})\\
\mathrm{s.t.}\quad &{\rm{C5}},{\rm{C6}},\notag
\end{align}
\noindent where $F(\bold{x},\bold{y},\bold{t})$ and $G(\bold{x},\bold{y},\bold{t})$ are given by:\vspace*{-2mm}
\begin{align}\label{eqn:P2_DCF}
 &F(\bold{x},\bold{y},\bold{t}) =  - \lambda N\sum\limits_{k = 1}^{K}{R^k_{{\rm{min}}}} - \eta N\sum\limits_{j = 1}^{{N_{{\rm{NF}}}}} {Q_{{\rm{NF}}}^{j2}} \notag\\
 &+ (1 + \lambda)\sum\limits_{n = 1}^N {\sum\limits_{k = 1}^K {\sum\limits_{i = 1}^{{N_\mathrm{F}}} {s_k^i[n]{{\log }_2}\left( {{t_k}[n] + {\gamma _0}{\cal P}} \right)} } },\quad\quad\;\;
\end{align}
\begin{align}\label{eqn:P2_DCG}
&G(\bold{x},\bold{y},\bold{t}) =\left( 1 + \lambda \right)\sum\limits_{n = 1}^N {\sum\limits_{k = 1}^K {\sum\limits_{i = 1}^{{N_\mathrm{F}}} {s_k^i[n]{{\log }_2}{t_k}[n]} } } \notag\\
&- \eta \sum\limits_{n = 1}^N \hspace{-0.05in} {\sum\limits_{j = 1}^{{N_{{\rm{NF}}}}} {{{(x[n] - x_{{\rm{NF}}}^j)}^2}} }  - \eta \sum\limits_{n = 1}^N \hspace{-0.05in} {\sum\limits_{j = 1}^{{N_{{\rm{NF}}}}} {{{(y[n] - y_{{\rm{NF}}}^j)}^2}} }.
\end{align}

Note that $F(\bold{x},\bold{y},\bold{t})$ and $G(\bold{x},\bold{y},\bold{t})$ are differentiable concave functions with respect to (w.r.t.) $\bold{x}$, $\bold{y}$, and $\bold{t}$.
Thus, for any feasible point $( {\bold{x}^{\left( l \right)}},{\bold{y}^{\left( l \right)}},{\bold{t}^{\left( l \right)}} )$, we can define the global upperestimator for $G(\bold{x},\bold{y},\bold{t})$ based on its first order Taylor's expansion at $( {\bold{x}^{\left( l \right)}},{\bold{y}^{\left( l \right)}},{\bold{t}^{\left( l \right)}} )$ as follows:\vspace*{-2mm}
\begin{align}\label{eqn:inequality of problem 2}
&G(\bold{x},\bold{y},\bold{t}) \hspace{-0.02in} \le \hspace{-0.02in} G(\hspace{-0.02in}{\bold{x}^{\left( l \right)}}\hspace{-0.02in},\hspace{-0.02in}{\bold{y}^{\left( l \right)}}\hspace{-0.02in},\hspace{-0.02in}{\bold{t}^{\left( l \right)}}\hspace{-0.02in}) + {\nabla _\bold{x}}G{(\hspace{-0.02in}{\bold{x}^{\left( l \right)}}\hspace{-0.02in},\hspace{-0.02in}{\bold{y}^{\left( l \right)}}\hspace{-0.02in},\hspace{-0.02in}{\bold{t}^{\left( l \right)}}\hspace{-0.02in})^T}(\bold{x} - {\bold{x}^{\left( l \right)}})\notag\\
& \hspace{-0.02in} + \hspace{-0.02in} {\nabla _\bold{y}}G{(\hspace{-0.02in}{\bold{x}^{\left( l \right)}}\hspace{-0.02in},\hspace{-0.02in}{\bold{y}^{\left( l \right)}}\hspace{-0.02in},\hspace{-0.02in}{\bold{t}^{\left( l \right)}}\hspace{-0.02in})^T}(\bold{y}\hspace{-0.04in} - \hspace{-0.04in} {\bold{y}^{\left( l \right)}}) \hspace{-0.02in} + \hspace{-0.02in} {\nabla _\bold{t}}G{(\hspace{-0.02in}{\bold{x}^{\left( l \right)}}\hspace{-0.02in},\hspace{-0.02in}{\bold{y}^{\left( l \right)}}\hspace{-0.02in},\hspace{-0.02in}{\bold{t}^{\left( l \right)}}\hspace{-0.02in})^T}(\bold{t} \hspace{-0.04in} - \hspace{-0.04in} {\bold{t}^{\left( l \right)}}),
\end{align}
\noindent where ${\nabla_\bold{x}}G( {\bold{x}^{\left( l \right)}},{\bold{y}^{\left( l \right)}},{\bold{t}^{\left( l \right)}} )$, ${\nabla_\bold{y}}G( {\bold{y}^{\left( l \right)}},{\bold{y}^{\left( l \right)}},{\bold{t}^{\left( l \right)}} )$, and ${\nabla_\bold{t}}G( {\bold{t}^{\left( l \right)}},{\bold{y}^{\left( l \right)}},{\bold{t}^{\left( l \right)}} )$ denote the gradient vectors of $G( {\bold{x}},{\bold{y}},{\bold{t}} )$  at $( {\bold{x}^{\left( l \right)}},{\bold{y}^{\left( l \right)}},{\bold{t}^{\left( l \right)}} )$.
Moreover, the right hand side of \eqref{eqn:inequality of problem 2} is an affine function.
Thus, we can acquire a lower bound for the optimal value of problem \eqref{eqn:P2_DCD} by solving the following concave maximization problem:

\begin{table}[t]\label{table:algorithm 2}\vspace*{-2mm}
  \begin{algorithm} [H]                
  \renewcommand\thealgorithm{2}
  \caption{Proposed Relay Trajectory Optimization}
  \label{algorithm:algorithm 2}
   \begin{algorithmic} [1]
   \STATE Initialize the maximum number of iterations $L_{\max}$, iteration index $l=0$, penalty factors $\lambda$ and $\eta$, the UAV's trajectory as $\bold x^{(0)}$, $\bold y^{(0)}$, and $\bold t^{(0)}$.

   \REPEAT
   \STATE Solve \eqref{eqn:P2_final} for a given $\bold x^{(l)}$, $\bold y^{(l)}$, $\bold t^{(l)}$ and obtain the intermediate optimal solution as $\{\bold x,\bold y,\bold t \}$.
   \STATE
   Set $l = l+1$ and $\bold x^{(l)}=\bold x$, $\bold y^{(l)}=\bold y$, and $\bold t^{(l)}=\bold t$.

   \UNTIL convergence or $l=L_{\max}$.

   \STATE
   $\bold x^{*} = \bold x^{(l)}$, $\bold y^{*} = \bold y^{(l)}$, and $\bold t^{*} = \bold t^{(l)}$.

   \end{algorithmic}
  \end{algorithm}
  \vspace*{-9.5mm}
\end{table}

\begin{align}\label{eqn:P2_final}
&\underset{\bold{x},\bold{y},\bold{t}}{\maxo}  F\hspace{-0.02in} (\hspace{-0.02in} \bold{x}, \hspace{-0.02in} \bold{y}, \hspace{-0.02in} \bold{t}\hspace{-0.02in} )
\hspace{-0.04in} - \hspace{-0.04in} G(\hspace{-0.025in}{\bold{x}^{\left(\hspace{-0.007in} l \hspace{-0.007in} \right)}}\hspace{-0.025in},\hspace{-0.025in}{\bold{y}^{\left(\hspace{-0.007in} l \hspace{-0.007in} \right)}}\hspace{-0.025in},\hspace{-0.025in}{\bold{t}^{\left(\hspace{-0.007in} l \hspace{-0.007in} \right)}}\hspace{-0.025in})
\hspace{-0.04in} - \hspace{-0.04in} {\nabla\hspace{-0.025in} _\bold{x}}\hspace{-0.025in} G{(\hspace{-0.025in}{\bold{x}^{\left(\hspace{-0.007in} l \hspace{-0.007in} \right)}}\hspace{-0.025in},\hspace{-0.025in}{\bold{y}^{\left(\hspace{-0.007in} l \hspace{-0.007in} \right)}}\hspace{-0.025in},\hspace{-0.025in}{\bold{t}^{\left(\hspace{-0.007in} l \hspace{-0.007in} \right)}}\hspace{-0.025in})^T} %
\hspace{-0.03in} (\bold{x} \hspace{-0.04in} - \hspace{-0.04in} {\bold{x}^{\left(\hspace{-0.007in} l \hspace{-0.007in} \right)}} ) )\notag\\
& \hspace{-0.04in} - \hspace{-0.04in} {\nabla _\bold{y}}G{(\hspace{-0.02in} {\bold{x}^{\left(\hspace{-0.007in} l \hspace{-0.007in} \right)}}\hspace{-0.02in}, \hspace{-0.02in} {\bold{y}^{\left(\hspace{-0.007in} l \hspace{-0.007in} \right)}}\hspace{-0.02in}, \hspace{-0.02in} {\bold{t}^{\left(\hspace{-0.007in} l \hspace{-0.007in} \right)}})^T}
\hspace{-0.02in}(\bold{y} \hspace{-0.04in} - \hspace{-0.04in} {\bold{y}^{\left(\hspace{-0.007in} l \hspace{-0.007in} \right)}})
\hspace{-0.04in} - \hspace{-0.04in} {\nabla _\bold{t}}G{(\hspace{-0.02in} {\bold{x}^{\left(\hspace{-0.007in} l \hspace{-0.007in} \right)}}\hspace{-0.02in}, \hspace{-0.02in} {\bold{y}^{\left(\hspace{-0.007in} l \hspace{-0.007in} \right)}}\hspace{-0.02in}, \hspace{-0.02in} {\bold{t}^{\left(\hspace{-0.007in} l \hspace{-0.007in} \right)}})^T}
\hspace{-0.02in}(\bold{t} \hspace{-0.04in} - \hspace{-0.04in} {\bold{t}^{\left(\hspace{-0.007in} l \hspace{-0.007in} \right)}})\\[-4mm]
&\quad\quad{\rm{s.t.}}\quad  {\rm{C5}},{\rm{C6}},\notag\vspace*{-4mm}
\end{align}\vspace*{-4mm}
where\vspace*{-2mm}

\begin{align}\label{eqn:P2_final_x}
&{\nabla _x} G  ( {\bold{x}^{\left( l  \right)}},  {\bold{y}^{\left( l  \right)}}, {\bold{t}^{\left( l  \right)}}))^T
( \bold{x} - {\bold{x}^{\left( l  \right)}} )\notag\\
= &2\eta  \sum\limits_{n = 1}^N  {\sum\limits_{j = 1}^{{N_{{\rm{NF}}}}}
{({x^{\left( l  \right)}}[n]  -  x_{{\rm{NF}}}^j )} }
( x[n] - {x^{\left( l \right)}}[n]), \quad  \quad \quad \quad
\end{align}\vspace*{-2mm}
\begin{align}\label{eqn:P2_final_y}
&{\nabla _y}  G( {\bold{x}^{\left( l \right)}}, {\bold{y}^{\left( l  \right)}}, {\bold{t}^{\left( l  \right)}}))^T
( \bold{y}  -  {\bold{y}^{\left( l  \right)}} ) \notag\\
= &2\eta  \sum\limits_{n = 1}^N  {\sum\limits_{j = 1}^{{N_{{\rm{NF}}}}}
{({y^{\left( l  \right)}}[n] - y_{{\rm{NF}}}^j )} }
(y[n] - {y^{\left( l  \right)}}[n]), \quad  \quad \quad \quad
\end{align}\vspace*{-2mm}
\begin{align}\label{eqn:P2_final_t}
&{\nabla _t}G({\bold{x}^{\left( l \right)}},{\bold{y}^{\left( l \right)}},{\bold{t}^{\left( l \right)}}))^T(\bold{t} - {\bold{t}^{\left( l \right)}}) \notag\\
=&(1 \hspace{-0.02in} + \hspace{-0.02in} \lambda )\sum\limits_{n = 1}^N {\sum\limits_{k = 1}^K {\sum\limits_{i = 1}^{{N_\mathrm{F}}} {s_k^i[n]\frac{1}{{{t_k^{\left( l \right)}}[n]\ln 2}}} } }(t_k[n] \hspace{-0.025in} - \hspace{-0.025in} {t_k^{\left( l \right)}}[n]).
\end{align}

Now, the optimization problem in \eqref{eqn:P2_final} is a convex optimization problem and can be solved efficiently by standard convex problem solvers such as CVX \cite{Boyd}.
To tighten the obtained lower bound, we utilize an iterative algorithm to generate a sequence of feasible solutions $({\bold{x}^{\left( l + 1 \right)}},{\bold{y}^{\left( l + 1 \right)}},{\bold{t}^{\left( l + 1 \right)}} )$ successively, cf. {\bf Algorithm \ref{algorithm:algorithm 2}}.
The initial feasible solution $(\bold x^{(0)},\bold y^{(0)},\bold t^{(0)})$ is obtained by solving the convex optimization problem in \eqref{eqn:inequality of problem 2} with $F(\bold{x},\bold{y},\bold{t})$ as the objective function \cite{wei2016power}.
The intermediate solution from the last iteration will be used to update the problem in \eqref{eqn:P2_final} and it will generate a feasible solution for the next iteration.
The iterative procedure will stop either the changes of optimization variables are smaller than a predefined convergence tolerance or the number of iteration reaches its maximum.\vspace*{-1.5mm}

\subsection{Overall Algorithm}

In summary, the proposed algorithm solves the two subproblems \eqref{eqn:problem_1.1} and \eqref{eqn:subproblem 2.1} in an alternating manner. Since the objective value of \eqref{eqn:problem 1} with the solutions obtained by solving subproblems \eqref{eqn:problem_1.1} and \eqref{eqn:subproblem 2.1} is non-decreasing over iteration, and the optimal value of \eqref{eqn:problem 1} is finite, the solution obtained by the proposed algorithm is guaranteed to converge to a suboptimal solution \cite{Zeng_2016_Throughput_Maximization_for_UAV}, \cite{wei2016power}. The details of the proposed algorithm are summarized in {\bf Algorithm \ref{algorithm:algorithm 3}}.\vspace*{-1mm}

\begin{table}[t]\label{table:algorithm 3}
  \begin{algorithm} [H]
  \renewcommand\thealgorithm{3}
  \caption{Iterative Resource Allocation and Trajectory Optimization}

  \label{algorithm:algorithm 3}
   \begin{algorithmic} [1]
   \STATE Initialize the maximum number of iterations $L'_{\rm{max}}$, iteration index $l'=0$, and resource allocation policy as $\bold s^{(0)}$.

   \REPEAT
   \STATE For the fixed resource allocation $\bold s^{(l')}$, obtain the intermediate optimal trajectory $(\bold x , \bold y)$ using Algorithm 2.
   \STATE For the fixed UAV's trajectory $(\bold x , \bold y)$, find the intermediate optimal resource allocation $\bold s$ using Algorithm 1.
   \STATE Set $l' = l'+1$ and $\bold s^{(l')}=\bold s$, $(\bold x , \bold y) = (\bold x^{(l')}, \bold y^{(l')})$.
   \UNTIL convergence or iteration index reaches to the maximum number.
   \STATE $\bold s^{*} = \bold s^{(l')}$, $(\bold x^{*}, \bold y^{*}) = (\bold x^{(l')},  \bold y^{(l')})$.
   \end{algorithmic}
  \end{algorithm}
\end{table}

\section{Numerical Results}

In this section, we investigate the performance of the proposed UAV-enabled communication scheme through simulations.
All user nodes are placed on the ground within the area of $1 \times 1$ ${\rm{km}^2}$.
The communication bandwith is 2 MHz with carrier center frequency at 2 GHz, the number of subcarrier $N_\mathrm{F} = 16$, and the noise power on each subcarrier is $-100$ dBm with channel gain $\beta_0 = -50$ dB at the reference distance $d_0 = 1$ m.
Therefore, the reference SNR can be acquired as $\gamma_0 = 80$ $\rm{dB}$. The UAV's total flying time $T = 50$ s and its maximum flying speed $v_{\max} = 50$ m/s with a fixed altitude of $H = 100$ m.
Furthermore, we assume that the UAV's initial location and the final location in 2D area is $\{x[0],y[0]\} = (0,0)$, and $\{x[N],y[N]\} = (0,1000)$.
The radii of all the NFZs are the same with $Q_\mathrm{NF} = 150$ m.
For illustration, all the trajectories are sampled every second in our simulation.
%
\vspace*{-1mm}

In our simulation, we consider three specific UAV trajectory design benchmark schemes:
(a) \textit{Without} NFZ, which the UAV flies with the ignorance of NFZ; (b) \textit{Detour with Straight Path}, which ensures that more locations within the area can be covered \cite{Zeng_2017_Completion_Time_Minimization_in_UAV}; (c) \textit{Straight Trajectory}, which is a generally less power-consuming flying strategy \cite{Zeng_2017_Energy_Efficient_UAV}.\vspace*{-1mm}

\subsection{One NFZ with Single-User}

\vspace*{0mm}
\begin{figure}[t]\vspace*{-3mm}
  \centering
  \includegraphics[width=3.2 in]{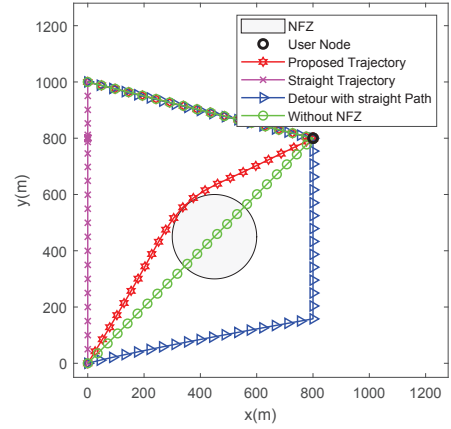}\vspace*{-4mm}
  \caption{Trajectories of a single-user case.}\vspace*{-4mm}
  \label{fig:Single Subcarrier Case}
\end{figure}
\vspace*{0mm}
\begin{figure}[t]
  \centering
  \includegraphics[width=2.7 in]{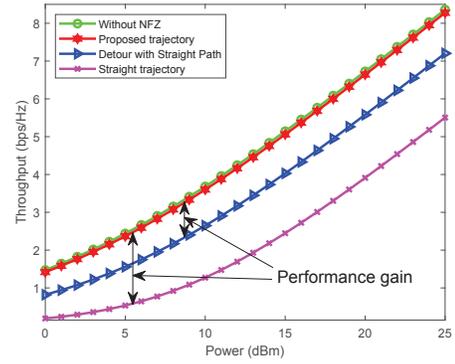}\vspace*{-3mm}
  \caption{Throughput versus transmission power per subcarrier for a single-user case.}\vspace*{-7mm}
  \label{fig:Throughput}
\end{figure}
Firstly, we consider a single-user located at $(800, 800, 0)$, and the minimum required data rate $R_{\rm{min}} = 3$ $\rm{bps/Hz}$ in each time slot $n$.
Besides, we assume that there is only one NFZ centered at $(450, 450)$ that blocks the UAV's straight path from the initial location to the user node.
We assume that the transmit power $\mathcal{P} = 10$ $\rm{dBm}$ on each subcarrier.
Then, the trajectories of the benchmark scheme and the proposed scheme are illustrated in Fig. \ref{fig:Single Subcarrier Case}.
It is observed that if there is no NFZ, the UAV will fly straightly to the user node and back to the final location after its hovering and communicating around the user.
In addition, the UAV always flies with the maximum speed $v_{\max} = 50$ m/s to acquire a shorter LoS link to the user as fast as possible.
Moreover, when the straight direction between the UAV and the user is blocked by the NFZ, the trajectory of the proposed scheme would take the shortest path by associated to the tangential line of the NFZ.

Now, we compare the system throughput achieved by the different trajectories.
Fig. \ref{fig:Throughput} illustrates the throughput in bps/Hz versus the transmit power per subcarrier.
It is first observed that the existence of the NFZ decreases the system throughput in general.
That is because that the UAV has to detour to avoid flying over the NFZ, that leaves less time of the UAV to reach closer to the user for hovering and effective communication.
Nonetheless, our proposed trajectory has only a slightly system performance degradation compared to the one without NFZ.
On the other hand, although the detour trajectory with straight path outperforms the straight trajectory, the throughput of which is much smaller than that of the proposed trajectory design.\vspace*{-1mm}

\subsection{Multi-NFZs with Multi-Users}

\begin{figure}[t]\vspace*{-1.5mm}
  \centering
  \includegraphics[width=3.2 in]{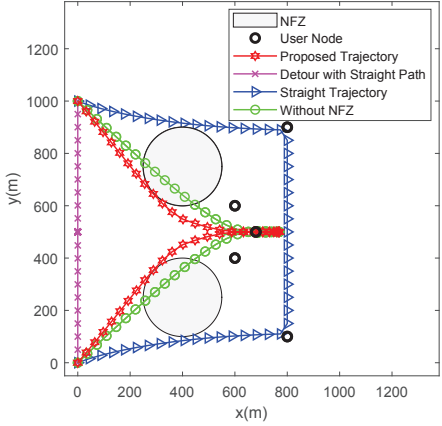}\vspace*{-4mm}
  \caption{Trajectories of multi-users case.}\vspace*{-4.5mm}
  \label{fig:Multi-NFZs} \vspace*{0mm}
\end{figure}\vspace*{0mm}

\begin{figure}[t]
  \centering
  \includegraphics[width=2.7 in]{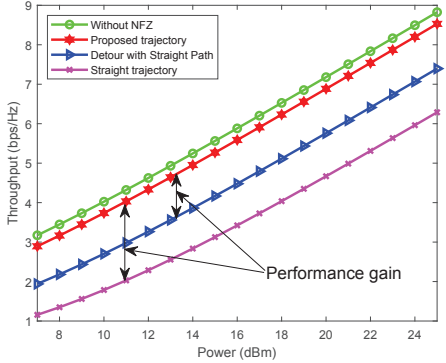}\vspace*{-3mm}
  \caption{Throughput versus transmission power per subcarrier for multi-users case.}\vspace*{-4mm}
  \label{fig:Five_Users_Throughput} \vspace*{0mm}
\end{figure}\vspace*{0mm}

Next, we consider a system with $K = 5$, $N_\mathrm{NF} = 2$, and all users have the same minimum required data rate $R_{\rm{min}} = 3$ $\rm{bps/Hz}$.
Also, the communication power on each subcarrier is set as $\mathcal{P} = 10$ $\rm{dBm}$.
As illustrated in Fig. \ref{fig:Multi-NFZs}, the UAV of the proposed scheme flies near to the centroid of the cluster formed by the users to achieve better channel gains by exploiting the high density of users.
Furthermore, the tangential trajectory still holds when the UAV flies to users from its initial location.

The throughput in bps/Hz versus the power per subcarrier in dBm achieved by various trajectories is illustrated in Fig. \ref{fig:Five_Users_Throughput}.
As can be observed, our proposed algorithm in multi-user case is also close to the system performance without NFZ.
However, due to more NFZs and users, the throughput gap between our proposed trajectory and the one without NFZ is slightly enlarged compared to the single-user case.
We also note that the joint trajectory and subcarrier allocation design becomes infeasible when the power on each subcarrier is lower than 7 dBm, whereas it is feasible in single-user case.
It is because that a higher transmit power is required to satisfy the minimum data rate requirements of more users.

\section{Conclusion}

This paper investigated a UAV-enabled OFDMA communication system providing communication services to ground users within an area which contains no-fly zones as required in some practical scenarios.
A joint trajectory and resource allocation algorithm was proposed to maximize the total throughput subject to UAV's mobility constraints, as well as the minimum instantaneous required data rate for each user.
Numerical results demonstrated that the proposed joint design algorithm, with consideration of the NFZs, can significantly increase the system throughput, compared to various benchmark schemes.

\section{Acknowledgement}
This work was supported by the Australia Research Council Discovery Project under Grant DP160104566, Linkage Project under Grant (LP 160100708), Discovery Early Career Researcher Award under Grant
DE170100137, and the China Scholarship Council.

\bibliographystyle{IEEEtran}

\begin{thebibliography}{99}
\providecommand{\url}[1]{#1}
\csname url@samestyle\endcsname
\providecommand{\newblock}{\relax}
\providecommand{\bibinfo}[2]{#2}
\providecommand{\BIBentrySTDinterwordspacing}{\spaceskip=0pt\relax}
\providecommand{\BIBentryALTinterwordstretchfactor}{4}
\providecommand{\BIBentryALTinterwordspacing}{\spaceskip=\fontdimen2\font plus
\BIBentryALTinterwordstretchfactor\fontdimen3\font minus
  \fontdimen4\font\relax}
\providecommand{\BIBforeignlanguage}[2]{{%
\expandafter\ifx\csname l@#1\endcsname\relax
\typeout{** WARNING: IEEEtran.bst: No hyphenation pattern has been}%
\typeout{** loaded for the language `#1'. Using the pattern for}%
\typeout{** the default language instead.}%
\else
\language=\csname l@#1\endcsname
\fi
#2}}
\providecommand{\BIBdecl}{\relax}
\BIBdecl

\bibitem{wong_2017_Key_Technologies_for_5G}
V.~W. Wong and L.-C. Wang, \emph{Key Technologies for {5G} Wireless
  Systems}.\hskip 1em plus 0.5em minus 0.4em\relax Cambridge University Press,
  2017.

\bibitem{Zeng_2016_Unmanned_Aerial_Vehicles_Magazine}
Y.~Zeng, R.~Zhang, and T.~J. Lim, ``Wireless communications with unmanned
  aerial vehicles: opportunities and challenges,'' \emph{IEEE Commun. Mag.},
  vol.~54, no.~5, pp. 36--42, May 2016.

\bibitem{Zeng_2016_Throughput_Maximization_for_UAV}
------, ``Throughput maximization for {UAV}-enabled mobile relaying systems,''
  \emph{IEEE Trans. Commun.}, vol.~64, no.~12, pp. 4983--4996, Dec. 2016.

\bibitem{Device_to_Device_Communications_Mozaffari}
M.~Mozaffari, W.~Saad, M.~Bennis, and M.~Debbah, ``Unmanned aerial vehicle with
  underlaid device-to-device communications: Performance and tradeoffs,''
  \emph{IEEE Trans. Wireless Commun.}, vol.~15, no.~6, pp. 3949--3963, Jun.
  2016.

\bibitem{Zeng_2017_Completion_Time_Minimization_in_UAV}
Y.~Zeng, X.~Xu, and R.~Zhang, ``Trajectory design for completion time
  minimization in {UAV}-enabled multicasting,'' \emph{IEEE Trans. Wireless
  Commun.}, vol.~17, no.~4, pp. 2233--2246, Apr. 2018.

\bibitem{Resource_Allocation_for_Solar_Powered}
Y.~Sun, D.~W.~K. Ng, D.~Xu, L.~Dai, and R.~Schober, ``Resource allocation for
  solar powered {UAV} communication systems,'' \emph{arXiv preprint
  arXiv:1801.07188}, 2018.

\bibitem{zhang2017cellular}
S.~Zhang, Y.~Zeng, and R.~Zhang, ``Cellular-enabled {UAV} communication:
  Trajectory optimization under connectivity constraint,'' \emph{arXiv preprint
  arXiv:1710.11619}, 2017.

\bibitem{Zhang_2017_Securing_UAV_Communications}
G.~Zhang, Q.~Wu, M.~Cui, and R.~Zhang, ``Securing {UAV} communications via
  trajectory optimization,'' \emph{arXiv preprint arXiv:1710.04389}, 2017.

\bibitem{Valavanis:2014:HUA:2692452}
K.~P. Valavanis and G.~J. Vachtsevanos, \emph{Handbook of Unmanned Aerial
  Vehicles}.\hskip 1em plus 0.5em minus 0.4em\relax Springer Publishing
  Company, Incorporated, 2014.

\bibitem{zhao_chen_yu_2017}
P.~Zhao, W.~Chen, and W.~Yu, ``Guidance law for intercepting target with
  multiple no-fly zone constraints,'' \emph{The Aeronautical Journal}, vol.
  121, no. 1244, p. 1479¨C1501, 2017.

\bibitem{A_Multiplier_Adjustment_Method}
M.~L. Fisher, R.~Jaikumar, and L.~N. Van~Wassenhove, ``A multiplier adjustment
  method for the generalized assignment problem,'' \emph{Management Science},
  vol.~32, no.~9, pp. 1095--1103, 1986.

\bibitem{Boyd}
S.~Boyd and L.~Vandenberghe, \emph{Convex optimization}.\hskip 1em plus 0.5em
  minus 0.4em\relax Cambridge University Press, 2004.

\bibitem{Optimal_Joint_Power_and_Subcarrier_Allocation}
Y.~Sun, D.~W.~K. Ng, Z.~Ding, and R.~Schober, ``Optimal joint power and
  subcarrier allocation for full-duplex multicarrier non-orthogonal multiple
  access systems,'' \emph{IEEE Trans. Commun.}, vol.~65, no.~3, pp. 1077--1091,
  Mar. 2017.

\bibitem{wei2016power}
Z.~Wei, D.~W.~K. Ng, J.~Yuan, and H.~M. Wang, ``Optimal resource allocation for
  power-efficient {MC-NOMA} with imperfect channel state information,''
  \emph{IEEE Trans. Commun.}, vol.~65, no.~9, pp. 3944--3961, Sep. 2017.

\bibitem{Zeng_2017_Energy_Efficient_UAV}
Y.~Zeng and R.~Zhang, ``Energy-efficient {UAV} communication with trajectory
  optimization,'' \emph{IEEE Trans. Wireless Commun.}, vol.~16, no.~6, pp.
  3747--3760, Jun. 2017.

\end{thebibliography}


\end{document}